\documentclass[a4paper, amsfonts, amssymb, amsmath, reprint, showkeys, nofootinbib, twoside, superscriptaddress, aps]{revtex4-1}
\usepackage{amsthm}
\usepackage[english]{babel}
\usepackage[utf8]{inputenc}
\usepackage[colorinlistoftodos, color=green!40, prependcaption]{todonotes}
\usepackage{mathtools}
\usepackage{physics}
\usepackage{xcolor}
\usepackage{graphicx}
\usepackage[left=23mm,right=13mm,top=35mm,columnsep=15pt]{geometry} 
\usepackage{adjustbox}
\usepackage{placeins}
\usepackage[T1]{fontenc}
\usepackage{lipsum}
\usepackage{csquotes}
\usepackage{tikz}
\usepackage{bm}
\usepackage{braket}
\usepackage{multirow}
\usepackage{qcircuit}
\usepackage{siunitx}
\usepackage{txfonts}
\usepackage{hyperref}
\hypersetup{
    colorlinks=true,
    linkcolor=blue,
    urlcolor=blue,
    citecolor=blue
    }
\usepackage[capitalise]{cleveref}

\newcommand{\R}{Ref.~}

\newcommand{\fig}{Fig.~}

\begin{document}

\title{Hardware-Efficient Randomized Compiling}

\author{Neelay Fruitwala}
    \thanks{These authors contributed equally to this work.}
    \affiliation{Accelerator Technology and Applied Physics Division, Lawrence Berkeley National Lab, Berkeley, CA 94720, USA}
\author{Akel Hashim}
    \thanks{These authors contributed equally to this work.}
    \affiliation{Department of Physics, University of California at Berkeley, Berkeley, CA 94720, USA}
    \affiliation{Applied Math and Computational Research Division, Lawrence Berkeley National Lab, Berkeley, CA 94720, USA}
\author{Abhi D.~Rajagopala}
    \affiliation{Applied Math and Computational Research Division, Lawrence Berkeley National Lab, Berkeley, CA 94720, USA}
\author{Yilun Xu}
    \affiliation{Accelerator Technology and Applied Physics Division, Lawrence Berkeley National Lab, Berkeley, CA 94720, USA}
\author{Jordan Hines}
    \affiliation{Department of Physics, University of California at Berkeley, Berkeley, CA 94720, USA}
\author{Ravi K.~Naik}
    \affiliation{Applied Math and Computational Research Division, Lawrence Berkeley National Lab, Berkeley, CA 94720, USA}
\author{Irfan Siddiqi}
    \affiliation{Department of Physics, University of California at Berkeley, Berkeley, CA 94720, USA}
    \affiliation{Applied Math and Computational Research Division, Lawrence Berkeley National Lab, Berkeley, CA 94720, USA}
    \affiliation{Materials Sciences Division, Lawrence Berkeley National Lab, Berkeley, CA 94720, USA}
\author{Katherine Klymko}
    \affiliation{National Energy Research Scientific Computing Center, Lawrence Berkeley National Lab, Berkeley, CA 94720, USA}
\author{Gang Huang}
    \thanks{Correspondence should be addressed to: \href{mailto:ghuang@lbl.gov }{ghuang@lbl.gov}.}
    \affiliation{Accelerator Technology and Applied Physics Division, Lawrence Berkeley National Lab, Berkeley, CA 94720, USA}
\author{Kasra Nowrouzi}
    \thanks{Correspondence should be addressed to: \href{mailto:kasra@berkeley.edu}{kasra@berkeley.edu}.}
    \affiliation{Applied Math and Computational Research Division, Lawrence Berkeley National Lab, Berkeley, CA 94720, USA}

\date{\today} 

\begin{abstract}
    Randomized compiling (RC) is an efficient method for tailoring arbitrary Markovian errors into stochastic Pauli channels. However, the standard procedure for implementing the protocol in software comes with a large experimental overhead --- namely, it scales linearly in the number of desired randomizations, each of which must be generated and measured independently. In this work, we introduce a hardware-efficient algorithm for performing RC on a cycle-by-cycle basis on the lowest level of our FPGA-based control hardware during the execution of a circuit. Importantly, this algorithm performs a different randomization per shot with zero runtime overhead beyond measuring a circuit without RC. We implement our algorithm using the \texttt{QubiC} control hardware, where we demonstrate significant reduction in the overall runtime of circuits implemented with RC, as well as a significantly lower variance in measured observables.
\end{abstract}


\maketitle

\section{Introduction} \label{sec:intro}

Randomized compiling (RC) \cite{wallman2016noise, hashim2021randomized} is a widely-used method for tailoring arbitrary Markovian errors acting on cycles in quantum circuits into stochastic Pauli noise channels. RC finds relevance in many areas of quantum computing, including error mitigation for NISQ applications \cite{ferracin2022efficiently}, benchmarking and characterization \cite{erhard2019characterizing, flammia2021averaged, proctor2022establishing, hines2023demonstrating, hashim2023benchmarking, carignan2023error}, phase estimation \cite{gu2023noise}, and quantum error correction \cite{jain2023improved, beale2023randomized}. Additionally, it is an efficient protocol, requiring few randomizations to achieve the desired level of error suppression \cite{hashim2021randomized} in a manner which is independent of system size \cite{goss2023extending}. Even so, the ideal limit for performing RC is to sample a different randomized circuit per experimental trial (i.e., ``shot'') --- which we term \emph{fully randomized compiling}, or FRC --- which can help minimize the variance of observables for many classes of circuits, such as randomized benchmarks \cite{granade2015accelerated, kwiatkowski2023optimized}. However, the standard method for performing RC has a high experimental overhead, as it requires generating many randomized logically-equivalent circuits in software prior to runtime, measuring each circuit independently, and combining the results in post-processing. Therefore, FRC remains impractical for applications which require more than a few tens or hundreds of shot. For this reason, improving the runtime performance of RC and related methods, such as Pauli frame randomization \cite{knill2004fault, kern2005quantum, ware2021experimental} and equivalent circuit averaging \cite{hashim2022optimized}, would be highly beneficial, enabling more randomizations to be measured in the same amount of clock time.

In this work, we introduce a hardware-efficient protocol --- which we term \emph{gateware-based RC} --- for implementing RC on the FPGA-based control hardware \texttt{QubiC} \cite{xu2021qubic, xu2023qubic}. Our protocol uses a low-level algorithm for performing RC on a cycle-by-cycle basis directly on the FPGA during the execution of a circuit. As long as single-qubit gate times are greater than 18 ns, this algorithm adds no additional runtime overhead beyond measuring a quantum circuit without RC, and can be used for any other protocol that utilizes Pauli twirling, such as cycle benchmarking \cite{erhard2019characterizing}, Pauli noise reconstruction \cite{flammia2020efficient, carignan2023error}, averaged circuit eigenvalue sampling \cite{flammia2021averaged, pelaez2024average}, mirror circuit fidelity estimation \cite{proctor2022establishing}, and many others. 

To demonstrate achievable performance gains, we provide detailed time profiling of our protocol and find that, in the FRC limit, gateware-based RC provides a $\sim250\times$ improvement in the total runtime compared to the standard software-based protocol. We additionally demonstrate its utility for cycle benchmarking, and show that we can measure the same gate fidelities as the software-based protocol, but with a significantly shorter runtime and smaller variance. Finally, we demonstrate that our protocol can improve the accuracy and lower the variance when estimating observables in circuits where systematic coherent errors are present.

\section{Gateware-based Randomized Compiling} \label{sec:hardware}

Traditionally, RC is performed entirely in software: Paulis are sampled uniformly at random before each two-qubit cycle in a circuit, the inverse operations are computed and inserted after each two-qubit gate cycle, and both the random Paulis and their inverses are compiled into the existing surrounding single-qubit gate cycles. This constitutes a single ``randomization'' of the original ``bare'' circuit, and this process is repeated many times to generate $n$ different randomizations of the bare circuit. Finally, each randomization is measured independently, and the results from all randomizations are combined in post-processing to generate an output distribution that is equivalent to a single quantum circuit. Therefore, performing RC imposes an $\mathcal{O}(n \times w \times d) $ overhead on the circuit compilation and upload time, where $n$ is the number of randomizations, and $w$ and $d$ are the circuit width and depth (in terms of two-qubit gate cycles), respectively.

Our hardware-efficient RC protocol performs random Pauli selection, commutation of these Paulis through the two-qubit gates, and single-qubit gate combination all on the control FPGA during circuit execution, eliminating the need to do these steps in software before runtime. Our method has zero additional experimental overhead for most single-qubit gate durations (gate times greater than 18 ns), and negligible compile time overhead, reducing the overall time complexity of compilation and upload to $\mathcal{O}(1 \times w \times d)$. We describe our protocol in detail below.

\subsection{Hardware-efficient Encoding Scheme}\label{sec:hwe_encoding}

In order to implement RC efficiently in the FPGA fabric, we encode the RC protocol using the following methods:

\subsubsection{Pauli Selection and Twirling}

At the beginning of each two-qubit gate cycle, a 2-bit-per-qubit pseudorandom number generator is used to draw a twirling gate from from the set of single-qubit Paulis, $\{I, X, Y, Z\}$. Here, we assume that all two-qubit gates are Clifford.\footnote{This could be extended to include non-Clifford two-qubit gates. But in this case, it would be necessary to restrict the set of Paulis to only those which commute with the gate. Thus, RC could only provide a partial (not full) twirl of the gate.} Therefore, propagating a pair of Pauli gates through a two-qubit gate yields another pair of Paulis. So, the inversion step can be represented as a 16-element lookup table, mapping the two-qubit Pauli $P_i \otimes P_j \in \{I, X, Y, Z\}^{\otimes 2}$ to another two-qubit Pauli $P_i' \otimes P_j' \in \{I, X, Y, Z\}^{\otimes 2}$, depending on which two-qubit Clifford gate(s) are used in the circuit.

\subsubsection{Single-qubit Gate Combination}\label{sec:rc_gate_comb}
Pauli twirling inserts Pauli gates around either side of each two-qubit gate cycle. If we assume that the circuit is structured into alternating cycles of single- and two-qubit gates, then the inserted Paulis will always be adjacent to another cycle of single-qubit gates. Therefore, any single qubit gate $U_3$ becomes $P_i U_3 P_{i- 1}'$, where $P_i$ is the current twirling gate and $P_{i - 1}'$ is the inversion gate from the previous cycle. 

Our gateware implementation assumes that each $U_3$ is decomposed into $Z(\phi_2) X_{90} Z(\phi_1) X_{90} Z(\phi_0)$ \cite{mckay2017efficient}, where $Z(\phi_i)$ are virtual-Z gates with arbitrary phases and $X_{90}$ is a $\pi/2$ gate about the $x$-axis. In this decomposition, Pauli operators can be absorbed into the virtual-Z phases: $P_i Z(\phi_2)X_{90} Z(\phi_1)X_{90}Z(\phi_0) P_{i-1}' = Z(\phi_2')X_{90} Z(\phi_1')X_{90}Z(\phi_0')$.  The new phases $\phi_n'$ are given by a function of the form: $\phi_n' = \pm \phi_n (\pm \pi)$, depending on the specific Pauli operators being absorbed. We implement absorption on the FPGA fabric as a 64-element map (four possibilities each for $P_i$ and $P_{i-1}'$, and $n \in \{0, 1, 2\})$ to the set of functions $f(\phi_n) = \{\phi_n, -\phi_n, \pi - \phi_n, \pi + \phi_n\}$. 

\subsection{FPGA Implementation}

\begin{figure*}
    \centering
    \includegraphics[width=2\columnwidth]{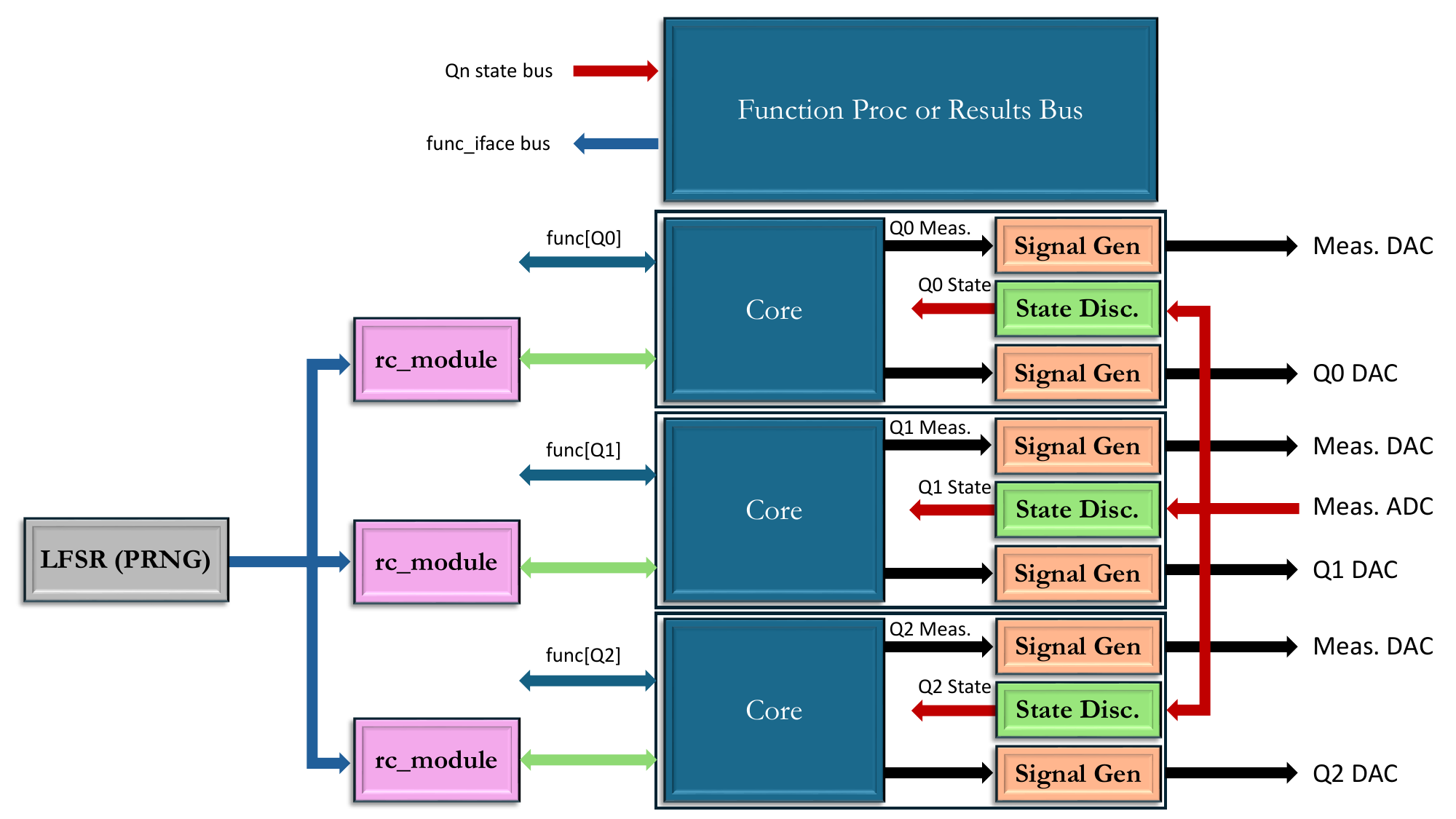}
    \caption{Block diagram of the gateware RC FPGA implementation and integration with \texttt{QubiC}. In the \texttt{QubiC} distributed architecture, each core is responsible for control and measurement of a single qubit. We connect every core to a dedicated \texttt{rc\_module}, which is responsible for inversion Pauli resolution and single-qubit gate combination for that qubit. All \texttt{rc\_modules} are connected to a global pseudorandom number generator (PRNG) for sampling twirling gates. Using a global PRNG ensures that every \texttt{rc\_module} is aware of the selected twirling gate for all of the qubits, which is necessary for computing the inversion Paulis across each two-qubit gates.}
    \label{fig:hw_rc_diagram}    
\end{figure*}

Our implementation of the encoding described above is tightly integrated with the \texttt{QubiC 2.0} \cite{xu2023qubic} Distributed Processor Architecture \cite{fruitwala2024distributed}. A schematic is given in \fig\ref{fig:hw_rc_diagram}. Random number generation for selecting twirling gates is performed \textit{globally}, using a $2 \times N_{qubit}$ bit width LFSR (linear feedback shift register) \cite{george2007linear}, which draws from a uniform distribution of pseudorandom numbers every FPGA clock cycle. Pauli selection, propagation, and single-qubit gate combination are performed \textit{locally} on a per-qubit basis using a bank of dedicated \verb|rc_module| gateware blocks.

The RC protocol is integrated into a pulse sequence using two instructions: \verb|latch_rc_cycle| and \verb|rc_alu|, which extend the QubiC ISA (instruction set architecture) \cite{fruitwala2024distributed}. We describe these below.

\subsubsection{The Latch Instruction}

The \texttt{latch\_rc\_cycle} instruction will assert an active-high trigger signal on the \texttt{rc\_module} to indicate the beginning of a new gate cycle, which in general consists of a cycle of single-qubit gates followed by a cycle of two-qubit gates. The \texttt{rc\_module} will then latch in the current output of the LFSR as the new set of twirling gates, and cache the previous set of twirling gates for the purposes of computing the inversion operations.

This instruction is timing controlled: the trigger signal indicating a new cycle is asserted at a specific timestamp provided in the instruction. This is to allow for trigger signals to be synchronized across different processor cores, ensuring that qubits sharing a two-qubit gate during a given cycle also share a common set of twirling gates (i.e., the $2 N_{qubit}$ LFSR value is latched in during the same clock cycle). The provided timestamp is referenced to the processor core internal counter, which is also used for controlling the pulse trigger time. 

\subsubsection{The RC ALU Instruction}

The \texttt{rc\_alu} instruction is used to modify virtual-Z phase parameters for the purposes of single-qubit gate combination (section \ref{sec:rc_gate_comb}). An initial phase is provided to the \texttt{rc\_module}, along with metadata indicating the previous gate in the two-qubit gate cycle (e.g., CNOT, CZ, or identity) and qubit pair (for computing the inversion operations for a two-qubit gate), as well as the location of the provided virtual-Z phase in the $Z(\phi_2)X_{90} Z(\phi_1)X_{90}Z(\phi_0)$ $U_3$ gate. The \texttt{rc\_module} will then determine the appropriate inversion Paulis from the provided metadata, and return a modified phase to the processor core. This modified phase will then be provided to the processor core ALU (arithmetic logic unit), where it can be saved to a register or used to increment a phase accumulator.

Properly utilizing the \verb|rc_alu| instruction requires that phase tracking is performed in hardware --- i.e. a processor core register is used to track the accumulated phase provided by the virtual-Z gates in the pulse sequence. This is straightforward to implement, and is a native feature of the \texttt{QubiC} gateware and compiler stack.

\begin{figure*}[t]
    \centering
    \includegraphics[width=2\columnwidth]{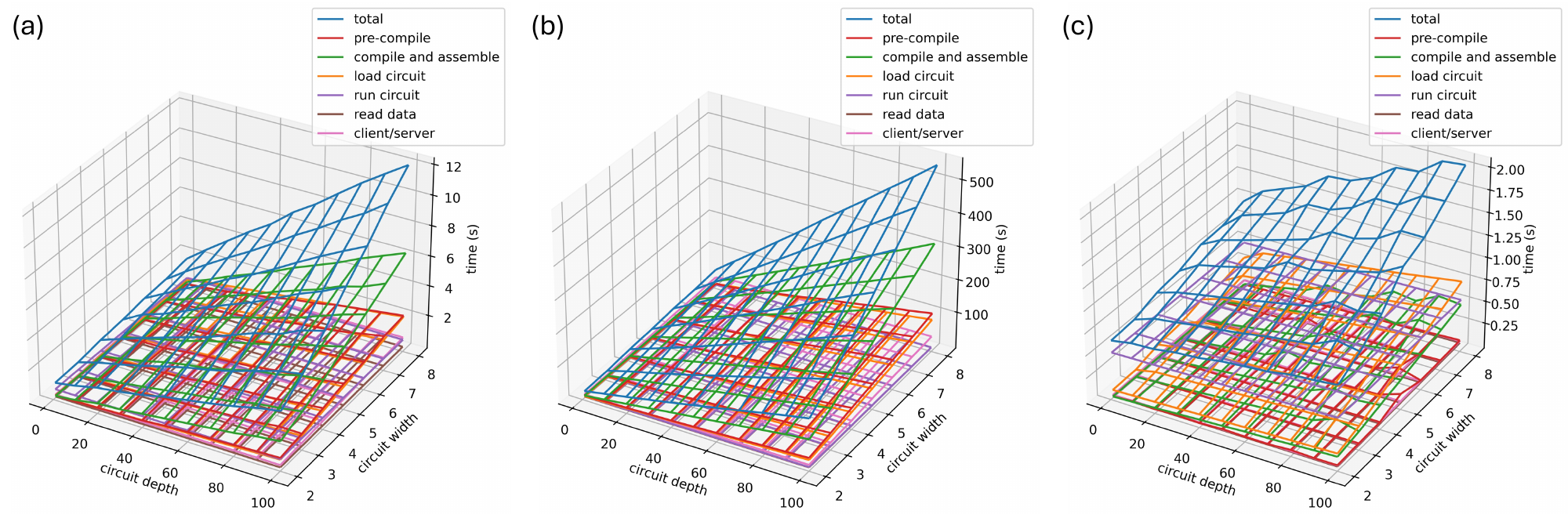}
    \caption{\textbf{Time Profiling.}
    We compare a breakdown of the execution time for performing RC on circuits of various widths (2 -- 8 qubits) and depths (1 -- 100 cycles of two-qubit gates) using the software-based and gateware-based RC protocols. For the software protocol, we implement RC using (a) 20 and (b) 1000 different randomizations over a total of 1000 shots. (c) The gateware-based protocol by default implements a different randomization per shot, for a total of 1000 randomizations over 1000 shots. We observe linear growth in the total execution time as a function of circuit depth and width for both versions of the protocol. At 1000 randomizations gateware RC provides a $\sim250\times$ speedup over software RC, and additionally provides a $\sim6\times$ speedup over the 20-randomization software case.
    }
    \label{fig:time}
\end{figure*}

\subsubsection{Timing Properties}

In the current implementation, the \verb|latch_rc_cycle| and \verb|rc_alu| instructions execute in 6 ns and 12 ns, respectively. These instructions can execute \textit{during} the playing of a control pulse. So, in practice, there is zero additional overhead for executing these RC instructions when single and two qubit gate times are greater than 18 ns. However, these timings are not a hard limit on how fast the RC instructions can be performed, nor have they been fully optimized; future work will explore ways to speedup these times or add specialized instruction pipelining to the distributed processor core.



\subsection{Compiler Integration}

The \texttt{QubiC} Intermediate Representation (\texttt{QubiC-IR}) \cite{fruitwala2024distributed, qubicirref} is a domain-specific intermediate representation for writing quantum programs targeted to \texttt{QubiC} hardware. \texttt{QubiC-IR} supports a variety of abstraction levels (e.g., both native gate and pulse-level control), and is designed to interface with standard software tools (such as \texttt{OpenQASM}) or programmed directly by the user.

As with the \texttt{QubiC} ISA, we extend \texttt{QubiC-IR} with two high-level instructions to implement gateware RC. The \texttt{LatchRcCycle} instruction is a higher-level version of the corresponding assembly instruction, and marks the beginning of each new single-qubit gate cycle. Unlike the assembly instruction, it accepts a list of qubits (or named frequencies) that are participating in the current cycle. The \texttt{QubiC} compiler will then schedule a \texttt{latch\_rc\_cycle} instruction on the relevant core(s) and ensure that the trigger times are synchronized across them. The \texttt{RcVirtualZ} instruction extends the normal \texttt{VirtualZ} instruction by adding the metadata fields required by the \texttt{rc\_alu} instruction. \texttt{RcVirtualZ} instructions are resolved into \texttt{rc\_alu} in a manner analogous to standard hardware-parameterized virtual-Z gates \cite{fruitwala2024distributed}.

\begin{figure*}[t]
    \centering
    \includegraphics[width=2\columnwidth]{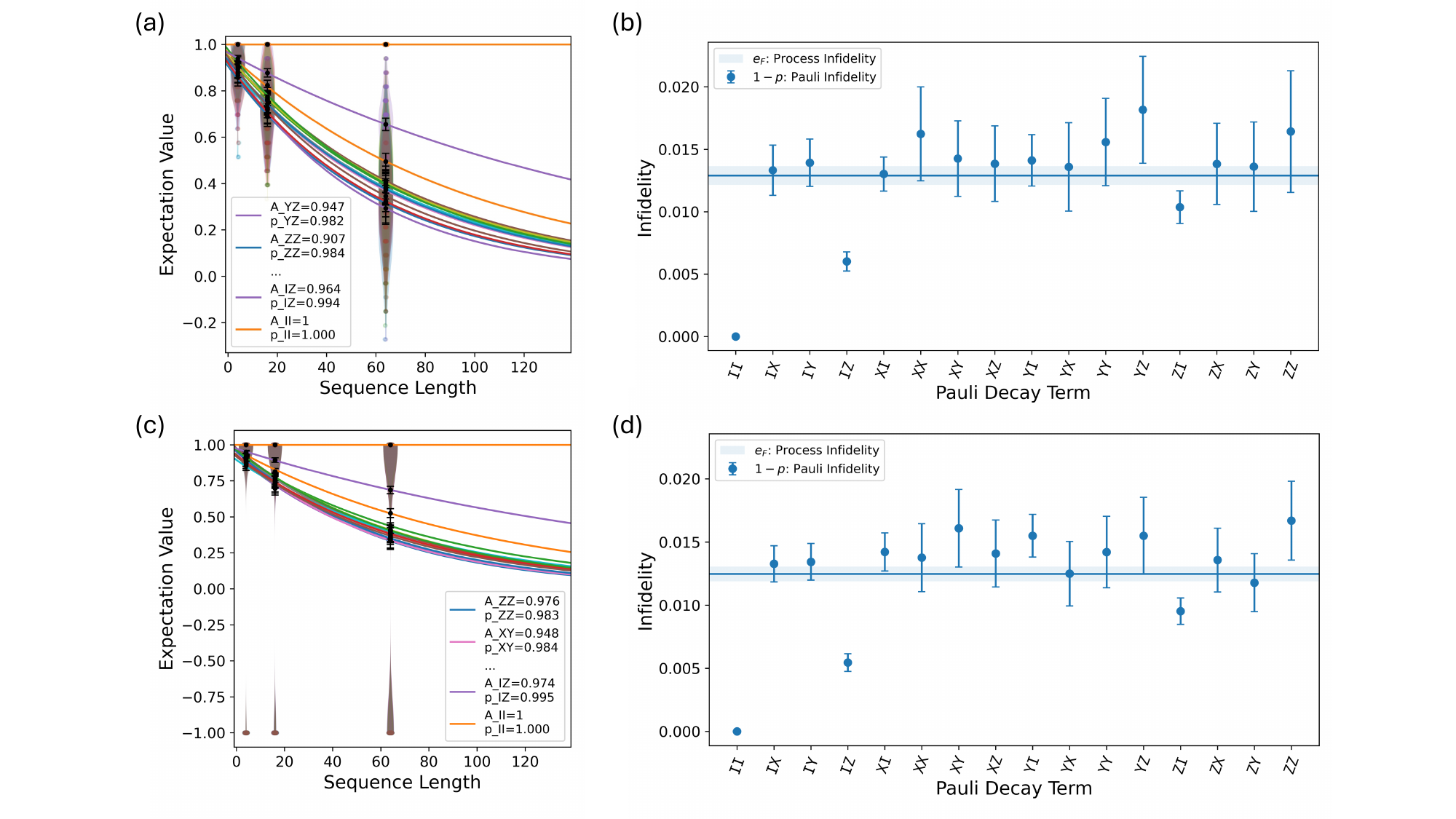}
    \caption{\textbf{Cycle Benchmarking.}
    (a) Exponential decays and (b) Pauli infidelities for software-based CB performed on a two-qubit CZ gate. (c) Exponential decays and (d) Pauli infidelities for gateware-based CB performed on the same gate. For (a) and (c), each Pauli decay is fit to an independent exponential function, with $A_P$ and $p_P$ denoting the SPAM (state-preparation and measurement) constant and the exponential fit parameter for each Pauli basis state $P$, respectively. The circular data points are the results of individual circuits, and the violin plots depict the distribution of results for each Pauli decay at each circuit depth; the bimodal distribution of results for individual circuits in (c) is expected behavior, since the expectation value of a circuit measured just once will either be $+1$ or $-1$. For (b) and (d), the average of the Pauli infidelities gives the process infidelity (horizontal line) of the dressed CZ gate; the opaque region behind the process infidelity denotes the uncertainty of the estimate. The process infidelities of both experiments are equal (up to uncertainty). We observe that the uncertainties of the individual Pauli infidelities for gateware-based CB are generally smaller than the corresponding uncertainties for the software-based results, as is expected in the FRC limit for the same total number of shots.
    }
    \label{fig:cb}
\end{figure*}

In addition to the above instructions, we also provide compiler directives to apply gateware RC automatically to selected portions of a quantum circuit. The compiler will then transform \texttt{VirtualZ} instructions into \texttt{RcVirtualZ} instructions with the metadata fields filled in, and add \texttt{LatchRcCycle} instructions to the beginning of every gate cycle. Using these directives requires that the circuit is described at the native-gate level.
\section{Time Profiling} \label{sec:time}

To benchmark the runtime performance of gateware-based RC against the standard software-based protocol, we profile the total execution time of each method on random circuits at varying depths (from 1 -- 100, defined by the number of two-qubit gate cycles) and widths (2 -- 8 qubits), shown in \fig\ref{fig:time}. Each circuit was measured 1000 times, split evenly across the total number of randomizations. For the software RC protocol, we measured both a typical number of randomizations (20 randomizations, 50 shots/each) \cite{hashim2021randomized}, as well as the FRC limit (1000 randomizations, 1 shot/each). We measure gateware RC only in the FRC limit, since the protocol will always randomize shot-to-shot. We use \texttt{True-Q} \cite{trueq} for generating the RC circuits for software RC.

We divide the total execution time into the following five fields:
\begin{itemize}
    \item \textbf{pre-compile}: this includes performing RC in software (for the standard protocol) and \texttt{True-Q} to \texttt{QubiC} transpilation.
    \item \textbf{compile and assemble}: compilation of the circuit(s) from high-level \texttt{QubiC-IR} to machine code, performed on a desktop computer.
    \item \textbf{load circuit}: loading compiled program, waveform memory, and any parameters into the FPGA fabric.\footnote{\textbf{load circuit} and \textbf{get data} \textit{only} include communication between the RFSoC PS (processing system) and PL (programmable logic), \textit{not} client/server communication over RPC (remote procedure call); in the current implementation, PS/PL communication is the dominant overhead.}
    \item \textbf{run circuit}: quantum measurement time.
    \item \textbf{get data}: retrieving measurement data from FPGA block RAM.
    \item \textbf{client/server}: communication between the \texttt{QubiC} board and desktop computer over a local network. This includes sending compiled circuits to the board and receiving measurement data.
\end{itemize}

Comparing the execution time of gateware- and software-based RC in the limit of large circuit depth and width, we observe that gateware RC provides a $\sim250\times$ speedup over software RC in the FRC limit, and even provides a $\sim6\times$ speedup over software RC measured with the more typical 20 randomizations. For software RC, the main temporal bottleneck is \textbf{compile and assemble}, due to the need to compile many different circuits from \texttt{QubiC-IR} to the low-level machine code. For gateware RC, \textbf{load circuit} is the largest classical overhead term, followed closely by \textbf{compile and assemble}. The large \textbf{load circuit} term is due to the data transfer time between the PS (processing system) and PL (programmable logic) parts of the RFSoC (radiofrequency system-on-a-chip). This  overhead is significantly ($\sim$$80 \times$) larger than the achievable limit given the capabilities of the hardware. We will address this in a future upgrade to \texttt{QubiC}, which will further improve the speedup provided by gateware RC. 

\section{Validation, Benchmarking, and Characterization}\label{sec:benchmarking}

\begin{figure*}[t]
    \centering
    \includegraphics[width=2\columnwidth]{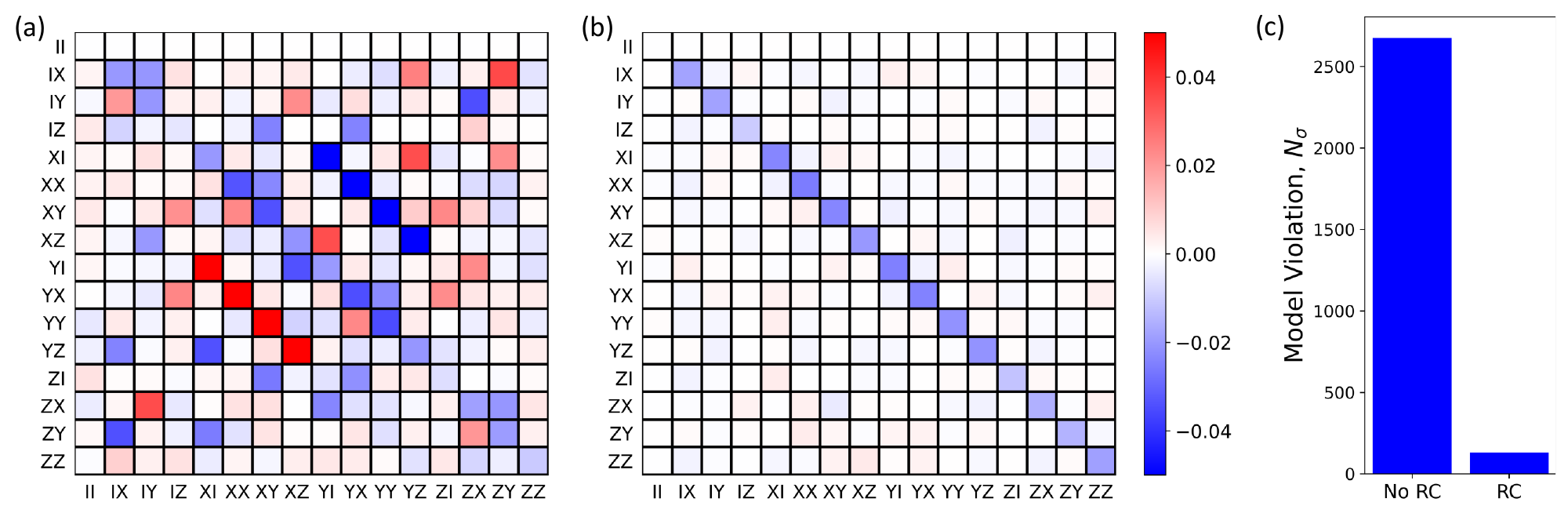}
    \caption{\textbf{Gate Set Tomography.}
    (a) PTM of the error generator for our two-qubit CZ gate measured without RC. 
    (b) PTM of the error generator for our two-qubit CZ gate measure with gateware RC. The colorbar denotes the magnitude of the errors in both PTMs; the smaller the magnitude, the closer that a given entry in the PTM matches the target gate. While many errors in the unital block of the PTM in (a) can be observed, the PTM in (b) contains mostly Pauli errors (all other off-diagonal terms are largely suppressed).
    (c) The model violation (i.e., goodness-of-fit) for the CPTP models fit to the GST data measured with and without RC. We observe an order of magnitude smaller model violation with gateware RC, suggesting that it provides a more trustworthy fit to the data. 
    }
    \label{fig:gst}
\end{figure*}

\subsection{Cycle Benchmarking}

The gateware RC protocol introduced in this work can be utilized in any circuit which takes advantage of Pauli twirling. In cycle benchmarking (CB) --- much like in RC --- circuits are composed of interleaved cycles of random Pauli gates with a gate (or cycle) of interest. In order to validate that our gateware RC protocol is implemented correctly, we benchmark a two-qubit CZ gate \cite{PhysRevLett.127.200502} using CB, where the randomization is performed in gateware, and compare the results obtained using the software-based protocol \cite{trueq}, shown in \fig\ref{fig:cb}. For software-based CB, we sample 30 different randomized circuits per circuit depth per Pauli decay term,\footnote{Here, each randomized circuit corresponds to a distinct randomization using RC, and each Pauli decay term corresponds to a different preparation and measurement basis.} for a total of 810 circuits across three different depths. We measure each circuit 33 times, for a total of $\sim 1$k shots per circuit depth per Pauli decay, with a total execution time of $\sim73.7$s. In \fig\ref{fig:cb}a, we plot the Pauli decays, and in \fig\ref{fig:cb}b, we plot the infidelities for each Pauli decay, from which we find a process infidelity of 1.29(4)\% for our (dressed) CZ gate.

Using our gateware RC protocol, it is only necessary to generate a single CB circuit per circuit depth per Pauli decay, for a total of 27 circuits, since RC is performed on a per-shot basis. We measure each circuit 1k times (giving 1k different randomizations per depth per Pauli decay), with a total execution time of $\sim16.8$s, a $\sim4.4\times$ reduction in runtime compared to software-based CB for the same amount of shot statistics. In \fig\ref{fig:cb}c, we plot the raw Pauli decays and observe that the expectation values for the individual circuit results are either $+1$ or $-1$, which is the expected behavior in the single-shot limit (i.e., a single outcome from any given circuit can be modeled by a Bernoulli distribution). In \fig\ref{fig:cb}d, we plot the Pauli decay infidelities, from which we find a process infidelity of 1.25(3)\% for our (dressed) CZ gate, which agrees with the software-based result up to the uncertainty. Moreover, we observe that the uncertainty of the individual Pauli infidelities are generally smaller than the corresponding Pauli infidelities from software-based CB, as would be expected in the FRC limit for the same total number of shots \cite{granade2015accelerated, kwiatkowski2023optimized}.

\subsection{Gate Set Tomography}

In the limit of perfect twirling (i.e., infinite randomizations), RC will tailor all \emph{Markovian} errors into stochastic Pauli channels. In the Pauli transfer matrix (PTM) superoperator formalism, Pauli channels are modeled by an error PTM with only diagonal entries; off-diagonal elements of a PTM arise either from non-unital errors (i.e., energy relaxation, such as $T_1$ decay) or coherent errors. To validate that our gateware RC protocol is tailoring noise as expected, we characterize our two-qubit CZ gate using gate set tomography (GST) \cite{blume2013robust, blume2017demonstration, nielsen2021gate} with and without RC \cite{hashim2023benchmarking}. In \fig\ref{fig:gst}, we plot the PTM of the error generator $\mathcal{L}$ \cite{blume2022taxonomy} of the bare and twirled CZ gate. The error generator $\mathcal{L}$ is roughly equivalent to the Linbladian superoperator that generates all Markovian errors (coherent, stochastic, non-unital, etc.) in the gate, which is modeled as $\Tilde{G} = e^\mathcal{L} G$, where $G$ is the ideal (noiseless) gate and $\Tilde{G}$ is the noisy gate. Using GST, we fit a two-qubit completely-positive and trace-preserving (CPTP) error model to the measured results for the CZ gate, from which we can extract $\mathcal{L}$. In \fig\ref{fig:gst}(a), we observe that the error generator for the bare gate has many non-zero entries in the unital (lower right-hand $15 \times 15$) block of the PTM, suggesting the presence of significant coherent and stochastic errors in the gate. In contrast, the error generator for the CZ gate measured with gateware RC applied to the GST sequences [\fig\ref{fig:gst}(b)] is much more sparse and diagonal, demonstrating the successful tailoring of noise in the gate. 

\begin{figure*}[t]
    \centering
    \includegraphics[width=2\columnwidth]{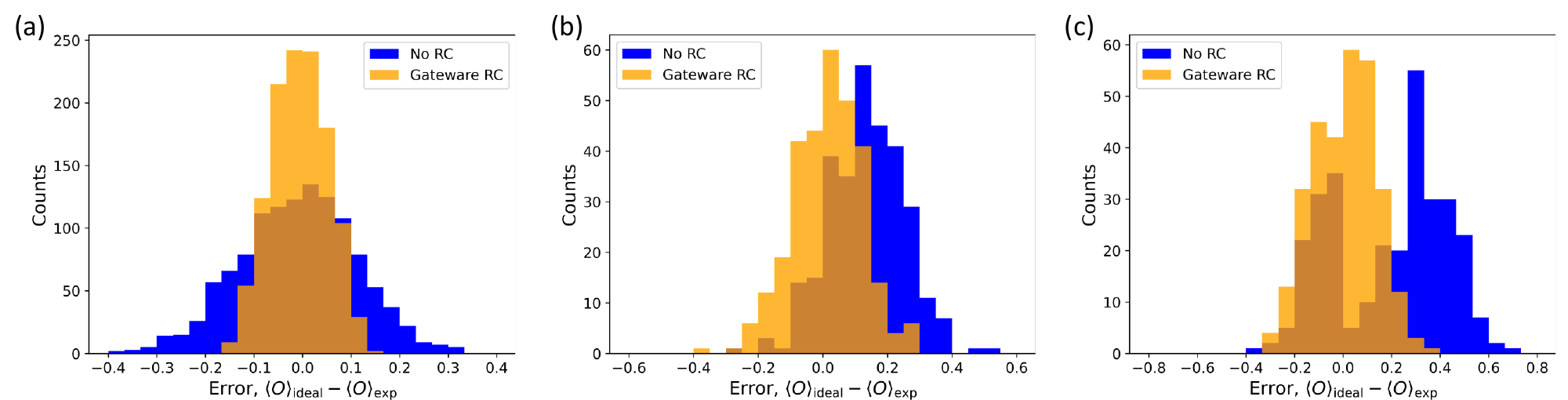}
    \caption{\textbf{Improving the Measurement of Observables.}
    (a) Histograms of the error in the observed expectation values for $\braket{IZ}$, $\braket{ZI}$, and $\braket{ZZ}$ for 400 different random two-qubit circuits measured with and without gateware RC (orange and blue, respectively). 
    (b) Histograms of the error in the expectation values for an average-case circuit calculated after subsampling 100 shots from the full 1000 shots 100 different times.
    (c) Histograms of the error in the expectation values for the worst-case circuit calculated after subsampling 100 shots from the full 1000 shots 100 different times.
    }
    \label{fig:variance}
\end{figure*}

GST also enables one to quantify how well a model can be fit to the observed data through a goodness-of-fit parameter called \emph{model violation}, denoted $N_\sigma$. Large model violation ($N_\sigma \gg 1$) signifies that there is strong statistical evidence in the data for errors not modeled by the GST fit. Since a complete (CPTP-constrained) two-qubit PTM model is capable of capturing any \emph{Markovian} CPTP process, errors that cannot be captured by the model are usually attributed to \emph{non-Markovianity} in the system. In \fig\ref{fig:gst}(c), we plot the model violation for the CPTP fits for the CZ gate measured with and without gateware RC.\footnote{The gate set also includes $I \otimes I$, $X_{\pi/2} \otimes I$, $Y_{\pi/2} \otimes I$, $I \otimes X_{\pi/2}$, and $I \otimes Y_{\pi/2}$. We focus primarily on the CZ gate, because this is the gate that we intentionally twirl in the GST sequences; the other gates are mostly needed for preparing fiducial states and measurements. However, it should be noted that model violation is computed for the \emph{entire} gate set, not just for the CZ gate.} We observe that the model violation without RC ($N_\sigma = 2672$) is an order of magnitude greater than the model violation with RC ($N_\sigma = 125$). This suggests that there is strong evidence for non-Markovian errors affecting the CZ gate measured without RC, but that these errors are largely suppressed by RC; a similar observation was made in \R\cite{hashim2023benchmarking}. Therefore, performing RC enables us to more reliably fit a complete CPTP model to our data, giving us a greater confidence about the errors we can expect to impact our gate.

\section{Improving the Measurement of Observables}\label{sec:variance}

Accurately estimating quantum observables is an important task for quantum computers. Performing quantum circuits and measuring expectation values of observables can be susceptible to both systematic errors and stochastic noise. Systematic errors, such as those resulting from coherent errors in a quantum circuit, can lead to the incorrect estimation of an observable, whereas stochastic noise will result in an increased variance of the observed quantity. Every quantum circuit is affected by both errors and noise to some extent, and thus the output of every circuit will have some bias; the more measurements we make, the more we learn about the bias. When we measure a circuit with RC, every randomization introduces a different bias in the results and, in the FRC limit, we learn minimal information about the bias of each randomization. By sampling from a maximal number of circuits with different biases (for a given number of shots), we can improve the accuracy \emph{and} reduce the variance of observables.

To demonstrate how we can improve the measurement of observables with our gateware RC protocol, we measure the expectation values of 400 different random two-qubit circuits with and without RC, similar to the experiment performed in \R\cite{ville2022leveraging}. We measure each circuit in the computational basis, compute the expectation value of $IZ$, $ZI$, and $ZZ$, and calculate the measured error in the expectation value relative to the (simulated) ideal outcome ($\braket{O}_\text{ideal} - \braket{O}_\text{exp}$). In \fig\ref{fig:variance}(a), we plot a histogram of the errors in all expectation values for all circuits measured with and without gateware RC; we denote these distributions by $X_\text{RC}$ and $X$, respectively. We observe that the distribution of both sets of results are centered around zero ($\overline{X} \approx \overline{X}_\text{RC} \approx 0$), but that the variance of $X_\text{RC}$ is an order of magnitude smaller than $X$ ($\text{Var}[X] \approx 1.2\times10^{-2}$ and $\text{Var}[X_\text{RC}] \approx 3.3\times10^{-3}$). In fact, in 71.4\% of all computed expectation values, the error in expectation value is smaller when measured with our gateware RC protocol than when measured without RC.

To explore the impact of gateware RC on individual circuits, we choose a circuit with an average performance and plot a histogram of the errors in the expectation values after subsampling 100 shots (from the full 1000 shots, where each shot is sampled at random without replacement) 100 different times, shown in \fig\ref{fig:variance}(b). This represents the distribution of expected errors in the observables if the circuit had only been measured 100 times. In this average-case example, we observe that $\text{Var}[X] \approx 1.4\times10^{-2}$ and $\text{Var}[X_\text{RC}] \approx 1.1\times10^{-2}$, showing only a modest reduction in variance. However, $\overline{X} \approx 0.14$ and $\overline{X}_\text{RC} \approx 0.016$, demonstrating the improved accuracy with which we can expect to measure expectation values with gateware RC. The large offset of $\overline{X}$ from zero is indicative of systematic errors in the circuit, which lead to a distinct bias in the results. Because we measure this circuit 1000 times, we learn a lot of information about this bias. In the gateware RC case, we learn nothing about the bias of each circuit, and instead sample from many different biased circuits, giving us a distribution that is, on average, centered around zero. To further highlight how pernicious the impact of systematic errors can be on measured observables, in \fig\ref{fig:variance}(c) we perform the same analysis for the circuit with the worst errors. For the bare circuit, we observe a bimodal distribution, suggesting that different expectation values suffer different errors, with the largest grouping of errors centered around $\overline{X} \approx 0.35$. On the other hand, for the gateware RC results, both the mean and variance are similar to the results obtained in (b), suggesting that we get a similar performance out of protocol, regardless of the degree to which systematic errors impact the results bare circuit. The results in \fig\ref{fig:variance} demonstrate that we can, in general, more accurately and reliably measure observables using gateware RC.
\section{Conclusions} \label{sec:conclusions}

RC is a useful protocol for simplifying the structure of noise in quantum circuits \cite{hashim2023benchmarking}. Moreover, it is particularly relevant to models for fault-tolerant quantum error correction (QEC), whose thresholds are typically computed assuming a stochastic (e.g., Pauli, dephasing, or depolarizing) noise model. However, the standard method of performing RC in software comes with a large experimental overhead, and will likely not be the modality in which RC is used for QEC. In this work, we introduce a hardware-efficient RC protocol that is implemented in the gateware of the open-source FPGA hardware \texttt{QubiC}. This protocol performs RC on a cycle-by-cycle basis \emph{during} the execution of a quantum circuit, and adds zero additional experimental overhead beyond measuring a circuit without RC as long as all gate times are longer than 18 ns. In this manner, one can simply turn ``on'' RC and run circuits as normal, without the need to measure multiple different sequences or combine results in post-processing. 

We demonstrate that our gateware RC protocol can reduce the overall runtime by more than two orders of magnitude compared to software-based RC when a different randomization is used per shot, and that the protocol can be used for any class of circuits that utilize Pauli twirling (e.g., cycle benchmarking). Additionally, we utilize GST to show that it efficiently tailors noise in gates, and provides us with more trustworthy models of gates. Finally, we demonstrate that the variance in observables can be minimized when sampling a different randomized circuit for each shot, which is particularly useful for obtaining estimates of gate fidelities using randomized benchmarks. Moreover, this extends to any circuit that is susceptible to systematic (i.e. coherent) errors in which one might want to compute expectation values. Looking forward, we believe our gateware RC protocol will be an invaluable tool for improving the results and runtime performance of many classes of circuits, such as those used for quantum error correction. We anticipate that future work will explore the various applications that can benefit from our hardware RC protocol.
\paragraph*{\textbf{\textup{Acknowledgements}}}\label{sec:acknowledgements}
This majority of this work was supported by the Laboratory Directed Research and Development Program of Lawrence Berkeley National Laboratory under U.S. Department of Energy Contract No.~DE-AC02-05CH11231. Y.X., R.K.N., I.S., and G.H.~acknowledge financial support (for the primary development of the \texttt{QubiC} hardware) from the U.S.~Department of Energy, Office of Science, Office of Advanced Scientific Computing Research Quantum Testbed Program under Contract No.~DE-AC02-05CH11231 and the Quantum Testbed Pathfinder Program.

A.H.~acknowledges fruitful discussions with Joel J.~Wallman, Arnaud Carignan-Dugas, Timothy J.~Proctor, and Pranav Gokhale.

\paragraph*{\textbf{\textup{Author Contributions}}}\label{sec:author_contributions}
N.F.~and A.H.~performed the experiment and analyzed the data. G.H.~and N.F.~designed the gateware RC protocol. A.D.R.~implemented the time profiling for \texttt{QubiC}. N.F., Y.X., and G.H.~developed the classical control hardware used in this work. J.H.~assisted in the design and analysis of the GST data. N.F., A.H., and G.H.~wrote the manuscript with input from all coauthors. R.K.N., I.S., K.K., G.H., and K.N.~supervised all work.

\paragraph*{\textbf{\textup{Competing Interests}}}\label{sec:competing_interests}
The gateware-based RC protocol presented in this work is protected under the U.S.~patent application no.~63/651,033 (patent pending), filed by Lawrence Berkeley National Laboratory on behalf of the following inventors: G.H., N.F., A.H., Y.X., A.D.R., R.K.N., K.N., and I.S.

\paragraph*{\textbf{\textup{Data Availability}}}\label{sec:data_availability}
All data are available from the corresponding authors upon reasonable request.

\bibliography{bibliography}

\begin{thebibliography}{34}%
\makeatletter
\providecommand \@ifxundefined [1]{%
 \@ifx{#1\undefined}
}%
\providecommand \@ifnum [1]{%
 \ifnum #1\expandafter \@firstoftwo
 \else \expandafter \@secondoftwo
 \fi
}%
\providecommand \@ifx [1]{%
 \ifx #1\expandafter \@firstoftwo
 \else \expandafter \@secondoftwo
 \fi
}%
\providecommand \natexlab [1]{#1}%
\providecommand \enquote  [1]{``#1''}%
\providecommand \bibnamefont  [1]{#1}%
\providecommand \bibfnamefont [1]{#1}%
\providecommand \citenamefont [1]{#1}%
\providecommand \href@noop [0]{\@secondoftwo}%
\providecommand \href [0]{\begingroup \@sanitize@url \@href}%
\providecommand \@href[1]{\@@startlink{#1}\@@href}%
\providecommand \@@href[1]{\endgroup#1\@@endlink}%
\providecommand \@sanitize@url [0]{\catcode `\\12\catcode `\$12\catcode `\&12\catcode `\#12\catcode `\^12\catcode `\_12\catcode `\%12\relax}%
\providecommand \@@startlink[1]{}%
\providecommand \@@endlink[0]{}%
\providecommand \url  [0]{\begingroup\@sanitize@url \@url }%
\providecommand \@url [1]{\endgroup\@href {#1}{\urlprefix }}%
\providecommand \urlprefix  [0]{URL }%
\providecommand \Eprint [0]{\href }%
\providecommand \doibase [0]{http://dx.doi.org/}%
\providecommand \selectlanguage [0]{\@gobble}%
\providecommand \bibinfo  [0]{\@secondoftwo}%
\providecommand \bibfield  [0]{\@secondoftwo}%
\providecommand \translation [1]{[#1]}%
\providecommand \BibitemOpen [0]{}%
\providecommand \bibitemStop [0]{}%
\providecommand \bibitemNoStop [0]{.\EOS\space}%
\providecommand \EOS [0]{\spacefactor3000\relax}%
\providecommand \BibitemShut  [1]{\csname bibitem#1\endcsname}%
\let\auto@bib@innerbib\@empty
\bibitem [{\citenamefont {Wallman}\ and\ \citenamefont {Emerson}(2016)}]{wallman2016noise}%
  \BibitemOpen
  \bibfield  {author} {\bibinfo {author} {\bibfnamefont {J.~J.}\ \bibnamefont {Wallman}}\ and\ \bibinfo {author} {\bibfnamefont {J.}~\bibnamefont {Emerson}},\ }\href {\doibase 10.1103/PhysRevA.94.052325} {\bibfield  {journal} {\bibinfo  {journal} {Phys. Rev. A}\ }\textbf {\bibinfo {volume} {94}},\ \bibinfo {pages} {052325} (\bibinfo {year} {2016})}\BibitemShut {NoStop}%
\bibitem [{\citenamefont {Hashim}\ \emph {et~al.}(2021)\citenamefont {Hashim}, \citenamefont {Naik}, \citenamefont {Morvan}, \citenamefont {Ville}, \citenamefont {Mitchell}, \citenamefont {Kreikebaum}, \citenamefont {Davis}, \citenamefont {Smith}, \citenamefont {Iancu}, \citenamefont {O'Brien}, \citenamefont {Hincks}, \citenamefont {Wallman}, \citenamefont {Emerson},\ and\ \citenamefont {Siddiqi}}]{hashim2021randomized}%
  \BibitemOpen
  \bibfield  {author} {\bibinfo {author} {\bibfnamefont {A.}~\bibnamefont {Hashim}}, \bibinfo {author} {\bibfnamefont {R.~K.}\ \bibnamefont {Naik}}, \bibinfo {author} {\bibfnamefont {A.}~\bibnamefont {Morvan}}, \bibinfo {author} {\bibfnamefont {J.-L.}\ \bibnamefont {Ville}}, \bibinfo {author} {\bibfnamefont {B.}~\bibnamefont {Mitchell}}, \bibinfo {author} {\bibfnamefont {J.~M.}\ \bibnamefont {Kreikebaum}}, \bibinfo {author} {\bibfnamefont {M.}~\bibnamefont {Davis}}, \bibinfo {author} {\bibfnamefont {E.}~\bibnamefont {Smith}}, \bibinfo {author} {\bibfnamefont {C.}~\bibnamefont {Iancu}}, \bibinfo {author} {\bibfnamefont {K.~P.}\ \bibnamefont {O'Brien}}, \bibinfo {author} {\bibfnamefont {I.}~\bibnamefont {Hincks}}, \bibinfo {author} {\bibfnamefont {J.~J.}\ \bibnamefont {Wallman}}, \bibinfo {author} {\bibfnamefont {J.}~\bibnamefont {Emerson}}, \ and\ \bibinfo {author} {\bibfnamefont {I.}~\bibnamefont {Siddiqi}},\ }\href {\doibase 10.1103/PhysRevX.11.041039} {\bibfield  {journal} {\bibinfo  {journal} {Phys. Rev.
  X}\ }\textbf {\bibinfo {volume} {11}},\ \bibinfo {pages} {041039} (\bibinfo {year} {2021})}\BibitemShut {NoStop}%
\bibitem [{\citenamefont {Ferracin}\ \emph {et~al.}(2022)\citenamefont {Ferracin}, \citenamefont {Hashim}, \citenamefont {Ville}, \citenamefont {Naik}, \citenamefont {Carignan-Dugas}, \citenamefont {Qassim}, \citenamefont {Morvan}, \citenamefont {Santiago}, \citenamefont {Siddiqi},\ and\ \citenamefont {Wallman}}]{ferracin2022efficiently}%
  \BibitemOpen
  \bibfield  {author} {\bibinfo {author} {\bibfnamefont {S.}~\bibnamefont {Ferracin}}, \bibinfo {author} {\bibfnamefont {A.}~\bibnamefont {Hashim}}, \bibinfo {author} {\bibfnamefont {J.-L.}\ \bibnamefont {Ville}}, \bibinfo {author} {\bibfnamefont {R.}~\bibnamefont {Naik}}, \bibinfo {author} {\bibfnamefont {A.}~\bibnamefont {Carignan-Dugas}}, \bibinfo {author} {\bibfnamefont {H.}~\bibnamefont {Qassim}}, \bibinfo {author} {\bibfnamefont {A.}~\bibnamefont {Morvan}}, \bibinfo {author} {\bibfnamefont {D.~I.}\ \bibnamefont {Santiago}}, \bibinfo {author} {\bibfnamefont {I.}~\bibnamefont {Siddiqi}}, \ and\ \bibinfo {author} {\bibfnamefont {J.~J.}\ \bibnamefont {Wallman}},\ }\href@noop {} {\bibfield  {journal} {\bibinfo  {journal} {arXiv preprint arXiv:2201.10672}\ } (\bibinfo {year} {2022})}\BibitemShut {NoStop}%
\bibitem [{\citenamefont {Erhard}\ \emph {et~al.}(2019)\citenamefont {Erhard}, \citenamefont {Wallman}, \citenamefont {Postler}, \citenamefont {Meth}, \citenamefont {Stricker}, \citenamefont {Martinez}, \citenamefont {Schindler}, \citenamefont {Monz}, \citenamefont {Emerson},\ and\ \citenamefont {Blatt}}]{erhard2019characterizing}%
  \BibitemOpen
  \bibfield  {author} {\bibinfo {author} {\bibfnamefont {A.}~\bibnamefont {Erhard}}, \bibinfo {author} {\bibfnamefont {J.~J.}\ \bibnamefont {Wallman}}, \bibinfo {author} {\bibfnamefont {L.}~\bibnamefont {Postler}}, \bibinfo {author} {\bibfnamefont {M.}~\bibnamefont {Meth}}, \bibinfo {author} {\bibfnamefont {R.}~\bibnamefont {Stricker}}, \bibinfo {author} {\bibfnamefont {E.~A.}\ \bibnamefont {Martinez}}, \bibinfo {author} {\bibfnamefont {P.}~\bibnamefont {Schindler}}, \bibinfo {author} {\bibfnamefont {T.}~\bibnamefont {Monz}}, \bibinfo {author} {\bibfnamefont {J.}~\bibnamefont {Emerson}}, \ and\ \bibinfo {author} {\bibfnamefont {R.}~\bibnamefont {Blatt}},\ }\href@noop {} {\bibfield  {journal} {\bibinfo  {journal} {Nature communications}\ }\textbf {\bibinfo {volume} {10}},\ \bibinfo {pages} {5347} (\bibinfo {year} {2019})}\BibitemShut {NoStop}%
\bibitem [{\citenamefont {Flammia}(2021)}]{flammia2021averaged}%
  \BibitemOpen
  \bibfield  {author} {\bibinfo {author} {\bibfnamefont {S.~T.}\ \bibnamefont {Flammia}},\ }\href@noop {} {\bibfield  {journal} {\bibinfo  {journal} {arXiv preprint arXiv:2108.05803}\ } (\bibinfo {year} {2021})}\BibitemShut {NoStop}%
\bibitem [{\citenamefont {Proctor}\ \emph {et~al.}(2022)\citenamefont {Proctor}, \citenamefont {Seritan}, \citenamefont {Nielsen}, \citenamefont {Rudinger}, \citenamefont {Young}, \citenamefont {Blume-Kohout},\ and\ \citenamefont {Sarovar}}]{proctor2022establishing}%
  \BibitemOpen
  \bibfield  {author} {\bibinfo {author} {\bibfnamefont {T.}~\bibnamefont {Proctor}}, \bibinfo {author} {\bibfnamefont {S.}~\bibnamefont {Seritan}}, \bibinfo {author} {\bibfnamefont {E.}~\bibnamefont {Nielsen}}, \bibinfo {author} {\bibfnamefont {K.}~\bibnamefont {Rudinger}}, \bibinfo {author} {\bibfnamefont {K.}~\bibnamefont {Young}}, \bibinfo {author} {\bibfnamefont {R.}~\bibnamefont {Blume-Kohout}}, \ and\ \bibinfo {author} {\bibfnamefont {M.}~\bibnamefont {Sarovar}},\ }\href@noop {} {\bibfield  {journal} {\bibinfo  {journal} {arXiv preprint arXiv:2204.07568}\ } (\bibinfo {year} {2022})}\BibitemShut {NoStop}%
\bibitem [{\citenamefont {Hines}\ \emph {et~al.}(2023)\citenamefont {Hines}, \citenamefont {Lu}, \citenamefont {Naik}, \citenamefont {Hashim}, \citenamefont {Ville}, \citenamefont {Mitchell}, \citenamefont {Kriekebaum}, \citenamefont {Santiago}, \citenamefont {Seritan}, \citenamefont {Nielsen} \emph {et~al.}}]{hines2023demonstrating}%
  \BibitemOpen
  \bibfield  {author} {\bibinfo {author} {\bibfnamefont {J.}~\bibnamefont {Hines}}, \bibinfo {author} {\bibfnamefont {M.}~\bibnamefont {Lu}}, \bibinfo {author} {\bibfnamefont {R.~K.}\ \bibnamefont {Naik}}, \bibinfo {author} {\bibfnamefont {A.}~\bibnamefont {Hashim}}, \bibinfo {author} {\bibfnamefont {J.-L.}\ \bibnamefont {Ville}}, \bibinfo {author} {\bibfnamefont {B.}~\bibnamefont {Mitchell}}, \bibinfo {author} {\bibfnamefont {J.~M.}\ \bibnamefont {Kriekebaum}}, \bibinfo {author} {\bibfnamefont {D.~I.}\ \bibnamefont {Santiago}}, \bibinfo {author} {\bibfnamefont {S.}~\bibnamefont {Seritan}}, \bibinfo {author} {\bibfnamefont {E.}~\bibnamefont {Nielsen}},  \emph {et~al.},\ }\href@noop {} {\bibfield  {journal} {\bibinfo  {journal} {Physical Review X}\ }\textbf {\bibinfo {volume} {13}},\ \bibinfo {pages} {041030} (\bibinfo {year} {2023})}\BibitemShut {NoStop}%
\bibitem [{\citenamefont {Hashim}\ \emph {et~al.}(2023)\citenamefont {Hashim}, \citenamefont {Seritan}, \citenamefont {Proctor}, \citenamefont {Rudinger}, \citenamefont {Goss}, \citenamefont {Naik}, \citenamefont {Kreikebaum}, \citenamefont {Santiago},\ and\ \citenamefont {Siddiqi}}]{hashim2023benchmarking}%
  \BibitemOpen
  \bibfield  {author} {\bibinfo {author} {\bibfnamefont {A.}~\bibnamefont {Hashim}}, \bibinfo {author} {\bibfnamefont {S.}~\bibnamefont {Seritan}}, \bibinfo {author} {\bibfnamefont {T.}~\bibnamefont {Proctor}}, \bibinfo {author} {\bibfnamefont {K.}~\bibnamefont {Rudinger}}, \bibinfo {author} {\bibfnamefont {N.}~\bibnamefont {Goss}}, \bibinfo {author} {\bibfnamefont {R.~K.}\ \bibnamefont {Naik}}, \bibinfo {author} {\bibfnamefont {J.~M.}\ \bibnamefont {Kreikebaum}}, \bibinfo {author} {\bibfnamefont {D.~I.}\ \bibnamefont {Santiago}}, \ and\ \bibinfo {author} {\bibfnamefont {I.}~\bibnamefont {Siddiqi}},\ }\href@noop {} {\bibfield  {journal} {\bibinfo  {journal} {npj Quantum Information}\ }\textbf {\bibinfo {volume} {9}},\ \bibinfo {pages} {109} (\bibinfo {year} {2023})}\BibitemShut {NoStop}%
\bibitem [{\citenamefont {Carignan-Dugas}\ \emph {et~al.}(2023)\citenamefont {Carignan-Dugas}, \citenamefont {Dahlen}, \citenamefont {Hincks}, \citenamefont {Ospadov}, \citenamefont {Beale}, \citenamefont {Ferracin}, \citenamefont {Skanes-Norman}, \citenamefont {Emerson},\ and\ \citenamefont {Wallman}}]{carignan2023error}%
  \BibitemOpen
  \bibfield  {author} {\bibinfo {author} {\bibfnamefont {A.}~\bibnamefont {Carignan-Dugas}}, \bibinfo {author} {\bibfnamefont {D.}~\bibnamefont {Dahlen}}, \bibinfo {author} {\bibfnamefont {I.}~\bibnamefont {Hincks}}, \bibinfo {author} {\bibfnamefont {E.}~\bibnamefont {Ospadov}}, \bibinfo {author} {\bibfnamefont {S.~J.}\ \bibnamefont {Beale}}, \bibinfo {author} {\bibfnamefont {S.}~\bibnamefont {Ferracin}}, \bibinfo {author} {\bibfnamefont {J.}~\bibnamefont {Skanes-Norman}}, \bibinfo {author} {\bibfnamefont {J.}~\bibnamefont {Emerson}}, \ and\ \bibinfo {author} {\bibfnamefont {J.~J.}\ \bibnamefont {Wallman}},\ }\href@noop {} {\bibfield  {journal} {\bibinfo  {journal} {arXiv preprint arXiv:2303.17714}\ } (\bibinfo {year} {2023})}\BibitemShut {NoStop}%
\bibitem [{\citenamefont {Gu}\ \emph {et~al.}(2023)\citenamefont {Gu}, \citenamefont {Ma}, \citenamefont {Forcellini},\ and\ \citenamefont {Liu}}]{gu2023noise}%
  \BibitemOpen
  \bibfield  {author} {\bibinfo {author} {\bibfnamefont {Y.}~\bibnamefont {Gu}}, \bibinfo {author} {\bibfnamefont {Y.}~\bibnamefont {Ma}}, \bibinfo {author} {\bibfnamefont {N.}~\bibnamefont {Forcellini}}, \ and\ \bibinfo {author} {\bibfnamefont {D.~E.}\ \bibnamefont {Liu}},\ }\href@noop {} {\bibfield  {journal} {\bibinfo  {journal} {Physical Review Letters}\ }\textbf {\bibinfo {volume} {130}},\ \bibinfo {pages} {250601} (\bibinfo {year} {2023})}\BibitemShut {NoStop}%
\bibitem [{\citenamefont {Jain}\ \emph {et~al.}(2023)\citenamefont {Jain}, \citenamefont {Iyer}, \citenamefont {Bartlett},\ and\ \citenamefont {Emerson}}]{jain2023improved}%
  \BibitemOpen
  \bibfield  {author} {\bibinfo {author} {\bibfnamefont {A.}~\bibnamefont {Jain}}, \bibinfo {author} {\bibfnamefont {P.}~\bibnamefont {Iyer}}, \bibinfo {author} {\bibfnamefont {S.~D.}\ \bibnamefont {Bartlett}}, \ and\ \bibinfo {author} {\bibfnamefont {J.}~\bibnamefont {Emerson}},\ }\href@noop {} {\bibfield  {journal} {\bibinfo  {journal} {Physical Review Research}\ }\textbf {\bibinfo {volume} {5}},\ \bibinfo {pages} {033049} (\bibinfo {year} {2023})}\BibitemShut {NoStop}%
\bibitem [{\citenamefont {Beale}\ and\ \citenamefont {Wallman}(2023)}]{beale2023randomized}%
  \BibitemOpen
  \bibfield  {author} {\bibinfo {author} {\bibfnamefont {S.~J.}\ \bibnamefont {Beale}}\ and\ \bibinfo {author} {\bibfnamefont {J.~J.}\ \bibnamefont {Wallman}},\ }\href@noop {} {\bibfield  {journal} {\bibinfo  {journal} {arXiv preprint arXiv:2306.13752}\ } (\bibinfo {year} {2023})}\BibitemShut {NoStop}%
\bibitem [{\citenamefont {Goss}\ \emph {et~al.}(2023)\citenamefont {Goss}, \citenamefont {Ferracin}, \citenamefont {Hashim}, \citenamefont {Carignan-Dugas}, \citenamefont {Kreikebaum}, \citenamefont {Naik}, \citenamefont {Santiago},\ and\ \citenamefont {Siddiqi}}]{goss2023extending}%
  \BibitemOpen
  \bibfield  {author} {\bibinfo {author} {\bibfnamefont {N.}~\bibnamefont {Goss}}, \bibinfo {author} {\bibfnamefont {S.}~\bibnamefont {Ferracin}}, \bibinfo {author} {\bibfnamefont {A.}~\bibnamefont {Hashim}}, \bibinfo {author} {\bibfnamefont {A.}~\bibnamefont {Carignan-Dugas}}, \bibinfo {author} {\bibfnamefont {J.~M.}\ \bibnamefont {Kreikebaum}}, \bibinfo {author} {\bibfnamefont {R.~K.}\ \bibnamefont {Naik}}, \bibinfo {author} {\bibfnamefont {D.~I.}\ \bibnamefont {Santiago}}, \ and\ \bibinfo {author} {\bibfnamefont {I.}~\bibnamefont {Siddiqi}},\ }\href@noop {} {\bibfield  {journal} {\bibinfo  {journal} {arXiv preprint arXiv:2305.16507}\ } (\bibinfo {year} {2023})}\BibitemShut {NoStop}%
\bibitem [{\citenamefont {Granade}\ \emph {et~al.}(2015)\citenamefont {Granade}, \citenamefont {Ferrie},\ and\ \citenamefont {Cory}}]{granade2015accelerated}%
  \BibitemOpen
  \bibfield  {author} {\bibinfo {author} {\bibfnamefont {C.}~\bibnamefont {Granade}}, \bibinfo {author} {\bibfnamefont {C.}~\bibnamefont {Ferrie}}, \ and\ \bibinfo {author} {\bibfnamefont {D.~G.}\ \bibnamefont {Cory}},\ }\href@noop {} {\bibfield  {journal} {\bibinfo  {journal} {New Journal of Physics}\ }\textbf {\bibinfo {volume} {17}},\ \bibinfo {pages} {013042} (\bibinfo {year} {2015})}\BibitemShut {NoStop}%
\bibitem [{\citenamefont {Kwiatkowski}\ \emph {et~al.}(2023)\citenamefont {Kwiatkowski}, \citenamefont {Stephenson}, \citenamefont {Knaack}, \citenamefont {Collopy}, \citenamefont {Bowers}, \citenamefont {Leibfried}, \citenamefont {Slichter}, \citenamefont {Glancy},\ and\ \citenamefont {Knill}}]{kwiatkowski2023optimized}%
  \BibitemOpen
  \bibfield  {author} {\bibinfo {author} {\bibfnamefont {A.}~\bibnamefont {Kwiatkowski}}, \bibinfo {author} {\bibfnamefont {L.~J.}\ \bibnamefont {Stephenson}}, \bibinfo {author} {\bibfnamefont {H.~M.}\ \bibnamefont {Knaack}}, \bibinfo {author} {\bibfnamefont {A.~L.}\ \bibnamefont {Collopy}}, \bibinfo {author} {\bibfnamefont {C.~M.}\ \bibnamefont {Bowers}}, \bibinfo {author} {\bibfnamefont {D.}~\bibnamefont {Leibfried}}, \bibinfo {author} {\bibfnamefont {D.~H.}\ \bibnamefont {Slichter}}, \bibinfo {author} {\bibfnamefont {S.}~\bibnamefont {Glancy}}, \ and\ \bibinfo {author} {\bibfnamefont {E.}~\bibnamefont {Knill}},\ }\href@noop {} {\bibfield  {journal} {\bibinfo  {journal} {arXiv preprint arXiv:2312.15836}\ } (\bibinfo {year} {2023})}\BibitemShut {NoStop}%
\bibitem [{\citenamefont {Knill}(2004)}]{knill2004fault}%
  \BibitemOpen
  \bibfield  {author} {\bibinfo {author} {\bibfnamefont {E.}~\bibnamefont {Knill}},\ }\href@noop {} {\bibfield  {journal} {\bibinfo  {journal} {arXiv preprint quant-ph/0404104}\ } (\bibinfo {year} {2004})}\BibitemShut {NoStop}%
\bibitem [{\citenamefont {Kern}\ \emph {et~al.}(2005)\citenamefont {Kern}, \citenamefont {Alber},\ and\ \citenamefont {Shepelyansky}}]{kern2005quantum}%
  \BibitemOpen
  \bibfield  {author} {\bibinfo {author} {\bibfnamefont {O.}~\bibnamefont {Kern}}, \bibinfo {author} {\bibfnamefont {G.}~\bibnamefont {Alber}}, \ and\ \bibinfo {author} {\bibfnamefont {D.~L.}\ \bibnamefont {Shepelyansky}},\ }\href@noop {} {\bibfield  {journal} {\bibinfo  {journal} {The European Physical Journal D-Atomic, Molecular, Optical and Plasma Physics}\ }\textbf {\bibinfo {volume} {32}},\ \bibinfo {pages} {153} (\bibinfo {year} {2005})}\BibitemShut {NoStop}%
\bibitem [{\citenamefont {Ware}\ \emph {et~al.}(2021)\citenamefont {Ware}, \citenamefont {Ribeill}, \citenamefont {Riste}, \citenamefont {Ryan}, \citenamefont {Johnson},\ and\ \citenamefont {Da~Silva}}]{ware2021experimental}%
  \BibitemOpen
  \bibfield  {author} {\bibinfo {author} {\bibfnamefont {M.}~\bibnamefont {Ware}}, \bibinfo {author} {\bibfnamefont {G.}~\bibnamefont {Ribeill}}, \bibinfo {author} {\bibfnamefont {D.}~\bibnamefont {Riste}}, \bibinfo {author} {\bibfnamefont {C.~A.}\ \bibnamefont {Ryan}}, \bibinfo {author} {\bibfnamefont {B.}~\bibnamefont {Johnson}}, \ and\ \bibinfo {author} {\bibfnamefont {M.~P.}\ \bibnamefont {Da~Silva}},\ }\href@noop {} {\bibfield  {journal} {\bibinfo  {journal} {Physical Review A}\ }\textbf {\bibinfo {volume} {103}},\ \bibinfo {pages} {042604} (\bibinfo {year} {2021})}\BibitemShut {NoStop}%
\bibitem [{\citenamefont {Hashim}\ \emph {et~al.}(2022)\citenamefont {Hashim}, \citenamefont {Rines}, \citenamefont {Omole}, \citenamefont {Naik}, \citenamefont {Kreikebaum}, \citenamefont {Santiago}, \citenamefont {Chong}, \citenamefont {Siddiqi},\ and\ \citenamefont {Gokhale}}]{hashim2022optimized}%
  \BibitemOpen
  \bibfield  {author} {\bibinfo {author} {\bibfnamefont {A.}~\bibnamefont {Hashim}}, \bibinfo {author} {\bibfnamefont {R.}~\bibnamefont {Rines}}, \bibinfo {author} {\bibfnamefont {V.}~\bibnamefont {Omole}}, \bibinfo {author} {\bibfnamefont {R.~K.}\ \bibnamefont {Naik}}, \bibinfo {author} {\bibfnamefont {J.~M.}\ \bibnamefont {Kreikebaum}}, \bibinfo {author} {\bibfnamefont {D.~I.}\ \bibnamefont {Santiago}}, \bibinfo {author} {\bibfnamefont {F.~T.}\ \bibnamefont {Chong}}, \bibinfo {author} {\bibfnamefont {I.}~\bibnamefont {Siddiqi}}, \ and\ \bibinfo {author} {\bibfnamefont {P.}~\bibnamefont {Gokhale}},\ }\href@noop {} {\bibfield  {journal} {\bibinfo  {journal} {Physical Review Research}\ }\textbf {\bibinfo {volume} {4}},\ \bibinfo {pages} {033028} (\bibinfo {year} {2022})}\BibitemShut {NoStop}%
\bibitem [{\citenamefont {Xu}\ \emph {et~al.}(2021)\citenamefont {Xu}, \citenamefont {Huang}, \citenamefont {Balewski}, \citenamefont {Naik}, \citenamefont {Morvan}, \citenamefont {Mitchell}, \citenamefont {Nowrouzi}, \citenamefont {Santiago},\ and\ \citenamefont {Siddiqi}}]{xu2021qubic}%
  \BibitemOpen
  \bibfield  {author} {\bibinfo {author} {\bibfnamefont {Y.}~\bibnamefont {Xu}}, \bibinfo {author} {\bibfnamefont {G.}~\bibnamefont {Huang}}, \bibinfo {author} {\bibfnamefont {J.}~\bibnamefont {Balewski}}, \bibinfo {author} {\bibfnamefont {R.}~\bibnamefont {Naik}}, \bibinfo {author} {\bibfnamefont {A.}~\bibnamefont {Morvan}}, \bibinfo {author} {\bibfnamefont {B.}~\bibnamefont {Mitchell}}, \bibinfo {author} {\bibfnamefont {K.}~\bibnamefont {Nowrouzi}}, \bibinfo {author} {\bibfnamefont {D.~I.}\ \bibnamefont {Santiago}}, \ and\ \bibinfo {author} {\bibfnamefont {I.}~\bibnamefont {Siddiqi}},\ }\href@noop {} {\bibfield  {journal} {\bibinfo  {journal} {IEEE Transactions on Quantum Engineering}\ }\textbf {\bibinfo {volume} {2}},\ \bibinfo {pages} {1} (\bibinfo {year} {2021})}\BibitemShut {NoStop}%
\bibitem [{\citenamefont {Xu}\ \emph {et~al.}(2023)\citenamefont {Xu}, \citenamefont {Huang}, \citenamefont {Fruitwala}, \citenamefont {Rajagopala}, \citenamefont {Naik}, \citenamefont {Nowrouzi}, \citenamefont {Santiago},\ and\ \citenamefont {Siddiqi}}]{xu2023qubic}%
  \BibitemOpen
  \bibfield  {author} {\bibinfo {author} {\bibfnamefont {Y.}~\bibnamefont {Xu}}, \bibinfo {author} {\bibfnamefont {G.}~\bibnamefont {Huang}}, \bibinfo {author} {\bibfnamefont {N.}~\bibnamefont {Fruitwala}}, \bibinfo {author} {\bibfnamefont {A.}~\bibnamefont {Rajagopala}}, \bibinfo {author} {\bibfnamefont {R.~K.}\ \bibnamefont {Naik}}, \bibinfo {author} {\bibfnamefont {K.}~\bibnamefont {Nowrouzi}}, \bibinfo {author} {\bibfnamefont {D.~I.}\ \bibnamefont {Santiago}}, \ and\ \bibinfo {author} {\bibfnamefont {I.}~\bibnamefont {Siddiqi}},\ }\href@noop {} {\bibfield  {journal} {\bibinfo  {journal} {arXiv preprint arXiv:2309.10333}\ } (\bibinfo {year} {2023})}\BibitemShut {NoStop}%
\bibitem [{\citenamefont {Flammia}\ and\ \citenamefont {Wallman}(2020)}]{flammia2020efficient}%
  \BibitemOpen
  \bibfield  {author} {\bibinfo {author} {\bibfnamefont {S.~T.}\ \bibnamefont {Flammia}}\ and\ \bibinfo {author} {\bibfnamefont {J.~J.}\ \bibnamefont {Wallman}},\ }\href@noop {} {\bibfield  {journal} {\bibinfo  {journal} {ACM Transactions on Quantum Computing}\ }\textbf {\bibinfo {volume} {1}},\ \bibinfo {pages} {1} (\bibinfo {year} {2020})}\BibitemShut {NoStop}%
\bibitem [{\citenamefont {Pelaez~Cisneros}\ \emph {et~al.}(2024)\citenamefont {Pelaez~Cisneros}, \citenamefont {Omole}, \citenamefont {Gokhale}, \citenamefont {Rines}, \citenamefont {Smith}, \citenamefont {Perlin},\ and\ \citenamefont {Hashim}}]{pelaez2024average}%
  \BibitemOpen
  \bibfield  {author} {\bibinfo {author} {\bibfnamefont {E.}~\bibnamefont {Pelaez~Cisneros}}, \bibinfo {author} {\bibfnamefont {V.}~\bibnamefont {Omole}}, \bibinfo {author} {\bibfnamefont {P.}~\bibnamefont {Gokhale}}, \bibinfo {author} {\bibfnamefont {R.}~\bibnamefont {Rines}}, \bibinfo {author} {\bibfnamefont {K.~N.}\ \bibnamefont {Smith}}, \bibinfo {author} {\bibfnamefont {M.~A.}\ \bibnamefont {Perlin}}, \ and\ \bibinfo {author} {\bibfnamefont {A.}~\bibnamefont {Hashim}},\ }\href@noop {} {\bibfield  {journal} {\bibinfo  {journal} {arXiv e-prints}\ ,\ \bibinfo {pages} {arXiv}} (\bibinfo {year} {2024})}\BibitemShut {NoStop}%
\bibitem [{\citenamefont {McKay}\ \emph {et~al.}(2017)\citenamefont {McKay}, \citenamefont {Wood}, \citenamefont {Sheldon}, \citenamefont {Chow},\ and\ \citenamefont {Gambetta}}]{mckay2017efficient}%
  \BibitemOpen
  \bibfield  {author} {\bibinfo {author} {\bibfnamefont {D.~C.}\ \bibnamefont {McKay}}, \bibinfo {author} {\bibfnamefont {C.~J.}\ \bibnamefont {Wood}}, \bibinfo {author} {\bibfnamefont {S.}~\bibnamefont {Sheldon}}, \bibinfo {author} {\bibfnamefont {J.~M.}\ \bibnamefont {Chow}}, \ and\ \bibinfo {author} {\bibfnamefont {J.~M.}\ \bibnamefont {Gambetta}},\ }\href@noop {} {\bibfield  {journal} {\bibinfo  {journal} {Physical Review A}\ }\textbf {\bibinfo {volume} {96}},\ \bibinfo {pages} {022330} (\bibinfo {year} {2017})}\BibitemShut {NoStop}%
\bibitem [{\citenamefont {Fruitwala}\ \emph {et~al.}(2024)\citenamefont {Fruitwala}, \citenamefont {Huang}, \citenamefont {Xu}, \citenamefont {Rajagopala}, \citenamefont {Hashim}, \citenamefont {Naik}, \citenamefont {Nowrouzi}, \citenamefont {Santiago},\ and\ \citenamefont {Siddiqi}}]{fruitwala2024distributed}%
  \BibitemOpen
  \bibfield  {author} {\bibinfo {author} {\bibfnamefont {N.}~\bibnamefont {Fruitwala}}, \bibinfo {author} {\bibfnamefont {G.}~\bibnamefont {Huang}}, \bibinfo {author} {\bibfnamefont {Y.}~\bibnamefont {Xu}}, \bibinfo {author} {\bibfnamefont {A.}~\bibnamefont {Rajagopala}}, \bibinfo {author} {\bibfnamefont {A.}~\bibnamefont {Hashim}}, \bibinfo {author} {\bibfnamefont {R.~K.}\ \bibnamefont {Naik}}, \bibinfo {author} {\bibfnamefont {K.}~\bibnamefont {Nowrouzi}}, \bibinfo {author} {\bibfnamefont {D.~I.}\ \bibnamefont {Santiago}}, \ and\ \bibinfo {author} {\bibfnamefont {I.}~\bibnamefont {Siddiqi}},\ }\href@noop {} {\bibfield  {journal} {\bibinfo  {journal} {arXiv preprint arXiv:2404.15260}\ } (\bibinfo {year} {2024})}\BibitemShut {NoStop}%
\bibitem [{\citenamefont {George}\ and\ \citenamefont {Alfke}(2007)}]{george2007linear}%
  \BibitemOpen
  \bibfield  {author} {\bibinfo {author} {\bibfnamefont {M.}~\bibnamefont {George}}\ and\ \bibinfo {author} {\bibfnamefont {P.}~\bibnamefont {Alfke}},\ }\href@noop {} {\bibfield  {journal} {\bibinfo  {journal} {Xilinx apprication note XAPP210}\ } (\bibinfo {year} {2007})}\BibitemShut {NoStop}%
\bibitem [{\citenamefont {Fruitwala}(2024)}]{qubicirref}%
  \BibitemOpen
  \bibfield  {author} {\bibinfo {author} {\bibfnamefont {N.}~\bibnamefont {Fruitwala}},\ }\href {https://lbl-qubic.gitlab.io/distributed_processor/} {\enquote {\bibinfo {title} {Qubic-ir languange reference},}\ } (\bibinfo {year} {2024})\BibitemShut {NoStop}%
\bibitem [{\citenamefont {Beale}\ \emph {et~al.}(2020)\citenamefont {Beale}, \citenamefont {Carignan-Dugas}, \citenamefont {Dahlen}, \citenamefont {Emerson}, \citenamefont {Hincks}, \citenamefont {Iyer}, \citenamefont {Jain}, \citenamefont {Hufnagel}, \citenamefont {Ospadov}, \citenamefont {Saunders}, \citenamefont {Stasiuk}, \citenamefont {Wallman},\ and\ \citenamefont {Winick}}]{trueq}%
  \BibitemOpen
  \bibfield  {author} {\bibinfo {author} {\bibfnamefont {S.~J.}\ \bibnamefont {Beale}}, \bibinfo {author} {\bibfnamefont {A.}~\bibnamefont {Carignan-Dugas}}, \bibinfo {author} {\bibfnamefont {D.}~\bibnamefont {Dahlen}}, \bibinfo {author} {\bibfnamefont {J.}~\bibnamefont {Emerson}}, \bibinfo {author} {\bibfnamefont {I.}~\bibnamefont {Hincks}}, \bibinfo {author} {\bibfnamefont {P.}~\bibnamefont {Iyer}}, \bibinfo {author} {\bibfnamefont {A.}~\bibnamefont {Jain}}, \bibinfo {author} {\bibfnamefont {D.}~\bibnamefont {Hufnagel}}, \bibinfo {author} {\bibfnamefont {E.}~\bibnamefont {Ospadov}}, \bibinfo {author} {\bibfnamefont {J.}~\bibnamefont {Saunders}}, \bibinfo {author} {\bibfnamefont {A.}~\bibnamefont {Stasiuk}}, \bibinfo {author} {\bibfnamefont {J.~J.}\ \bibnamefont {Wallman}}, \ and\ \bibinfo {author} {\bibfnamefont {A.}~\bibnamefont {Winick}},\ }\href {\doibase 10.5281/zenodo.3945250} {\enquote {\bibinfo {title} {True-q},}\ } (\bibinfo {year} {2020})\BibitemShut {NoStop}%
\bibitem [{\citenamefont {Mitchell}\ \emph {et~al.}(2021)\citenamefont {Mitchell}, \citenamefont {Naik}, \citenamefont {Morvan}, \citenamefont {Hashim}, \citenamefont {Kreikebaum}, \citenamefont {Marinelli}, \citenamefont {Lavrijsen}, \citenamefont {Nowrouzi}, \citenamefont {Santiago},\ and\ \citenamefont {Siddiqi}}]{PhysRevLett.127.200502}%
  \BibitemOpen
  \bibfield  {author} {\bibinfo {author} {\bibfnamefont {B.~K.}\ \bibnamefont {Mitchell}}, \bibinfo {author} {\bibfnamefont {R.~K.}\ \bibnamefont {Naik}}, \bibinfo {author} {\bibfnamefont {A.}~\bibnamefont {Morvan}}, \bibinfo {author} {\bibfnamefont {A.}~\bibnamefont {Hashim}}, \bibinfo {author} {\bibfnamefont {J.~M.}\ \bibnamefont {Kreikebaum}}, \bibinfo {author} {\bibfnamefont {B.}~\bibnamefont {Marinelli}}, \bibinfo {author} {\bibfnamefont {W.}~\bibnamefont {Lavrijsen}}, \bibinfo {author} {\bibfnamefont {K.}~\bibnamefont {Nowrouzi}}, \bibinfo {author} {\bibfnamefont {D.~I.}\ \bibnamefont {Santiago}}, \ and\ \bibinfo {author} {\bibfnamefont {I.}~\bibnamefont {Siddiqi}},\ }\href {\doibase 10.1103/PhysRevLett.127.200502} {\bibfield  {journal} {\bibinfo  {journal} {Phys. Rev. Lett.}\ }\textbf {\bibinfo {volume} {127}},\ \bibinfo {pages} {200502} (\bibinfo {year} {2021})}\BibitemShut {NoStop}%
\bibitem [{\citenamefont {Blume-Kohout}\ \emph {et~al.}(2013)\citenamefont {Blume-Kohout}, \citenamefont {Gamble}, \citenamefont {Nielsen}, \citenamefont {Mizrahi}, \citenamefont {Sterk},\ and\ \citenamefont {Maunz}}]{blume2013robust}%
  \BibitemOpen
  \bibfield  {author} {\bibinfo {author} {\bibfnamefont {R.}~\bibnamefont {Blume-Kohout}}, \bibinfo {author} {\bibfnamefont {J.~K.}\ \bibnamefont {Gamble}}, \bibinfo {author} {\bibfnamefont {E.}~\bibnamefont {Nielsen}}, \bibinfo {author} {\bibfnamefont {J.}~\bibnamefont {Mizrahi}}, \bibinfo {author} {\bibfnamefont {J.~D.}\ \bibnamefont {Sterk}}, \ and\ \bibinfo {author} {\bibfnamefont {P.}~\bibnamefont {Maunz}},\ }\href@noop {} {\bibfield  {journal} {\bibinfo  {journal} {arXiv preprint arXiv:1310.4492}\ } (\bibinfo {year} {2013})}\BibitemShut {NoStop}%
\bibitem [{\citenamefont {Blume-Kohout}\ \emph {et~al.}(2017)\citenamefont {Blume-Kohout}, \citenamefont {Gamble}, \citenamefont {Nielsen}, \citenamefont {Rudinger}, \citenamefont {Mizrahi}, \citenamefont {Fortier},\ and\ \citenamefont {Maunz}}]{blume2017demonstration}%
  \BibitemOpen
  \bibfield  {author} {\bibinfo {author} {\bibfnamefont {R.}~\bibnamefont {Blume-Kohout}}, \bibinfo {author} {\bibfnamefont {J.~K.}\ \bibnamefont {Gamble}}, \bibinfo {author} {\bibfnamefont {E.}~\bibnamefont {Nielsen}}, \bibinfo {author} {\bibfnamefont {K.}~\bibnamefont {Rudinger}}, \bibinfo {author} {\bibfnamefont {J.}~\bibnamefont {Mizrahi}}, \bibinfo {author} {\bibfnamefont {K.}~\bibnamefont {Fortier}}, \ and\ \bibinfo {author} {\bibfnamefont {P.}~\bibnamefont {Maunz}},\ }\href@noop {} {\bibfield  {journal} {\bibinfo  {journal} {Nature communications}\ }\textbf {\bibinfo {volume} {8}},\ \bibinfo {pages} {14485} (\bibinfo {year} {2017})}\BibitemShut {NoStop}%
\bibitem [{\citenamefont {Nielsen}\ \emph {et~al.}(2021)\citenamefont {Nielsen}, \citenamefont {Gamble}, \citenamefont {Rudinger}, \citenamefont {Scholten}, \citenamefont {Young},\ and\ \citenamefont {Blume-Kohout}}]{nielsen2021gate}%
  \BibitemOpen
  \bibfield  {author} {\bibinfo {author} {\bibfnamefont {E.}~\bibnamefont {Nielsen}}, \bibinfo {author} {\bibfnamefont {J.~K.}\ \bibnamefont {Gamble}}, \bibinfo {author} {\bibfnamefont {K.}~\bibnamefont {Rudinger}}, \bibinfo {author} {\bibfnamefont {T.}~\bibnamefont {Scholten}}, \bibinfo {author} {\bibfnamefont {K.}~\bibnamefont {Young}}, \ and\ \bibinfo {author} {\bibfnamefont {R.}~\bibnamefont {Blume-Kohout}},\ }\href@noop {} {\bibfield  {journal} {\bibinfo  {journal} {Quantum}\ }\textbf {\bibinfo {volume} {5}},\ \bibinfo {pages} {557} (\bibinfo {year} {2021})}\BibitemShut {NoStop}%
\bibitem [{\citenamefont {Blume-Kohout}\ \emph {et~al.}(2022)\citenamefont {Blume-Kohout}, \citenamefont {da~Silva}, \citenamefont {Nielsen}, \citenamefont {Proctor}, \citenamefont {Rudinger}, \citenamefont {Sarovar},\ and\ \citenamefont {Young}}]{blume2022taxonomy}%
  \BibitemOpen
  \bibfield  {author} {\bibinfo {author} {\bibfnamefont {R.}~\bibnamefont {Blume-Kohout}}, \bibinfo {author} {\bibfnamefont {M.~P.}\ \bibnamefont {da~Silva}}, \bibinfo {author} {\bibfnamefont {E.}~\bibnamefont {Nielsen}}, \bibinfo {author} {\bibfnamefont {T.}~\bibnamefont {Proctor}}, \bibinfo {author} {\bibfnamefont {K.}~\bibnamefont {Rudinger}}, \bibinfo {author} {\bibfnamefont {M.}~\bibnamefont {Sarovar}}, \ and\ \bibinfo {author} {\bibfnamefont {K.}~\bibnamefont {Young}},\ }\href@noop {} {\bibfield  {journal} {\bibinfo  {journal} {PRX Quantum}\ }\textbf {\bibinfo {volume} {3}},\ \bibinfo {pages} {020335} (\bibinfo {year} {2022})}\BibitemShut {NoStop}%
\bibitem [{\citenamefont {Ville}\ \emph {et~al.}(2022)\citenamefont {Ville}, \citenamefont {Morvan}, \citenamefont {Hashim}, \citenamefont {Naik}, \citenamefont {Lu}, \citenamefont {Mitchell}, \citenamefont {Kreikebaum}, \citenamefont {O'Brien}, \citenamefont {Wallman}, \citenamefont {Hincks}, \citenamefont {Emerson}, \citenamefont {Smith}, \citenamefont {Younis}, \citenamefont {Iancu}, \citenamefont {Santiago},\ and\ \citenamefont {Siddiqi}}]{ville2022leveraging}%
  \BibitemOpen
  \bibfield  {author} {\bibinfo {author} {\bibfnamefont {J.-L.}\ \bibnamefont {Ville}}, \bibinfo {author} {\bibfnamefont {A.}~\bibnamefont {Morvan}}, \bibinfo {author} {\bibfnamefont {A.}~\bibnamefont {Hashim}}, \bibinfo {author} {\bibfnamefont {R.~K.}\ \bibnamefont {Naik}}, \bibinfo {author} {\bibfnamefont {M.}~\bibnamefont {Lu}}, \bibinfo {author} {\bibfnamefont {B.}~\bibnamefont {Mitchell}}, \bibinfo {author} {\bibfnamefont {J.-M.}\ \bibnamefont {Kreikebaum}}, \bibinfo {author} {\bibfnamefont {K.~P.}\ \bibnamefont {O'Brien}}, \bibinfo {author} {\bibfnamefont {J.~J.}\ \bibnamefont {Wallman}}, \bibinfo {author} {\bibfnamefont {I.}~\bibnamefont {Hincks}}, \bibinfo {author} {\bibfnamefont {J.}~\bibnamefont {Emerson}}, \bibinfo {author} {\bibfnamefont {E.}~\bibnamefont {Smith}}, \bibinfo {author} {\bibfnamefont {E.}~\bibnamefont {Younis}}, \bibinfo {author} {\bibfnamefont {C.}~\bibnamefont {Iancu}}, \bibinfo {author} {\bibfnamefont {D.~I.}\ \bibnamefont {Santiago}}, \ and\ \bibinfo {author} {\bibfnamefont
  {I.}~\bibnamefont {Siddiqi}},\ }\href {\doibase 10.1103/PhysRevResearch.4.033140} {\bibfield  {journal} {\bibinfo  {journal} {Phys. Rev. Res.}\ }\textbf {\bibinfo {volume} {4}},\ \bibinfo {pages} {033140} (\bibinfo {year} {2022})}\BibitemShut {NoStop}%
\end{thebibliography}%

\clearpage
\appendix
\setcounter{table}{0}
\renewcommand{\thetable}{A\arabic{table}}

\setcounter{figure}{0}
\renewcommand{\thefigure}{A\arabic{figure}}

    

\end{document}